\documentclass[aps,showpacs, pra,amssymb, amsmath, twocolumn]{revtex4}
\usepackage{amsmath}
\usepackage{amsthm}
\usepackage{eucal}
\usepackage{amssymb}
\usepackage{mathrsfs}
\usepackage{graphicx}
\usepackage{amsmath,latexsym,amssymb}
\usepackage{graphicx}
\usepackage{amsmath}
\usepackage{latexsym}
\usepackage{amssymb}
\usepackage{bm}

\begin{document}

\title{Rabi oscillations of Morris-Shore transformed $N$-state systems by elliptically polarized ultrafast laser pulses}

\author{Hyosub Kim, Yunheung Song, Han-gyeol Lee,  and Jaewook Ahn}
\affiliation{Department of Physics, KAIST, Daejeon 305-701, Korea}

\date{\today}

\begin{abstract}
We present an experimental investigation of ultrafast-laser driven Rabi oscillations of atomic rubidium. Since the broadband spectrum of an ultrafast laser pulse simultaneously couples all the electronic hyperfine transitions between the excited and ground states, the complex excitation linkages involved with the D1 or D2 transition are energy degenerate. Here, by applying the Morris-Shore transformation, it is shown that this multi-state system is reduced to a set of independent two-state systems and dark states. In experiments performed by ultrafast laser interactions of atomic rubidium in the strong interaction regime, we demonstrate that the ultrafast dynamics of the considered multi-state system is governed by a sum of at most two decoupled Rabi oscillations when this system interacts with ultrafast laser pulses of any polarization state. We further show the implication of this result to possible controls of photo-electron polarizations.
\end{abstract}

\maketitle

\section{Introduction}
The Morris-Shore (MS) transformation~\cite{MorrisPRA1983} offers a theoretical framework to minimize the complexity in a variety of coherent excitation coupling linkages of a $N$-state quantum system~\cite{ShoreJMP2013}. Often used in modeling laser-induced excitation of atoms, the MS transformation replaces the complex system by a set of independent two-state systems, decoupled from each other, and unlinked states. For example, when a degenerate two-state atom possessing angular momenta $J$ and $J'$, respectively, for the ground and excited levels, interacts with a laser field, the excitation degeneracy in the magnetic sub-levels results in various coupling linkages amongst $N=2(J+J'+1)$ levels, characterized by polarization states of the laser field. However, a consequence of the MS-transformation decoupling is that the $N\times N$ Hamiltonian that describes the coherent excitation is reduced to a set of $2\times 2$ sub-Hamiltonians plus a diagonal matrix representing uncoupled dark or spectator states. So, the dynamics of such system can be described simply by a superposition of independent two-state dynamics, each of which undergoes a Rabi oscillation between a pair of new ground and excited states defined in a  factorized Hilbert sub-space. 

The procedure of the MS transformation can be summarized as follows~\cite{VitanovPRA2003}. Suppose an atomic system with $N_1$ ground levels and $N_2$ excited levels ($N=N_1+ N_2$) resonantly interacts with a laser field. Under an assumption of no damping, the interaction Hamiltonian is written in the rotating-wave approximation (RWA)~\cite{RWA,
Meystre,Boyd} as 
\begin{eqnarray}
H= 
\begin{pmatrix}
0 & V   \\ 
V^\dagger & 0  
\end{pmatrix},
\label{Int_H}
\end{eqnarray}
where $V$ is an $N_1 \times N_2$ coupling matrix between the ground and excited levels. Then, the MS transformation theory predicts that the coupling matrix $V$ is diagonalized through a MS transformation given by
\begin{equation}
H'=UHU^{\dagger},
\end{equation}
where the unitary transformation $U$ defined by
  \begin{equation}
  U=\begin{pmatrix}
  X & 0\\
  0 & Y
  \end{pmatrix}
  \label{U}.
  \end{equation}
Here the block matrices $X$ and $Y$ are the unitary transformations that diagonalize $VV^\dagger$ and $V^\dagger V$, respectively. The resulting Hamiltonian $H'$ is a direct sum of $M$ sets of two-state systems and $N-2M$ uncoupled single states ($2M \le \{N_1,N_2\}$), {\it i.e.,}
\begin{eqnarray}
H'&=&\hbar 
\begin{pmatrix}
0 & \Omega_1 \\
\Omega_1^* & 0
  \end{pmatrix} \oplus 
  \begin{pmatrix}
0 & \Omega_2 \\
\Omega_2^* & 0
  \end{pmatrix} \oplus \cdots  \oplus
    \begin{pmatrix}
0 & \Omega_M \\
\Omega_M^* & 0
  \end{pmatrix} \nonumber \\
  && \oplus 
  \begin{pmatrix}
h_1
  \end{pmatrix} \oplus
 \begin{pmatrix}
h_2
  \end{pmatrix}   \oplus \cdots  \oplus
   \begin{pmatrix}
h_{N-2M}
  \end{pmatrix},
  \end{eqnarray}
where $\Omega_i$'s are the Rabi frequencies, that are all different in general, of the two-state systems.

Various systems have been analyzed by the MS transformation~\cite{ShoreJMP2013}, which include two-state superposition systems~\cite{VitanovOC2000,VitanovJPB2000}, three-state $\Lambda$-linkage systems~\cite{VitanovARPC2001}, four-state diamond- and tripod-linkage systems~\cite{EberlyPRA1977, BialynickaPRA1977, ShorePRA1984, UnanyanOC1998,TheuerOE1999}, $N$-state $M$- and $W$-linkage systems~\cite{VitanovPRA2003}, to list a few. The MS transformation of three-state $\Lambda$-linkage systems is in particular interesting in regards to the laser technique called stimulated Raman adiabatic passage (STIRAP)~\cite{BergmannRMP1998}, where the systems undergo an adiabatic evolution between MS-transformed coupled two states, leaving the dark state never populated during the laser interaction. The result is a complete population transfer (CPT)~\cite{VitanovARPC2001,CPT,CPT2} from one ground state to the other, which is easy to understand in the context of the MS transformation.

In this paper, we describe an experimental study of MS-transformed $N$-state systems performed by ultrafast laser-induced coherent excitations. The systems under consideration are the complex coupling linkages of the D1 and D2 transitions of atomic rubidium ( $^{85}$Rb). For example, the D1 transitions from $|5S_{1/2}, F=2,3\rangle$ to $|5P_{1/2}, F'=2,3\rangle$ correspond to an $N=24$  system with $N_1=12$ and $N_2=12$. We first analyze the atomic rubidium systems in terms of the MS transformation to derive the Rabi frequencies for the D1 and D2 transitions induced by an elliptically-polarized light in Sec. II. After briefly describing the experimental procedure in Sec. III,  it is shown based on our experimental results that the system dynamics are all governed by a sum of at most two decoupled Rabi oscillations~\cite{RabiPR, RabiPR2} when these systems interact with ultrafast laser pulses of any polarization state. in Sec. IV. The implication of this result to possible controls of photo-electron polarizations is discussed before the conclusion in Sec. V.

\section{Theoretical Consideration}
\label{SecTheory}
\subsection{Linear and circular polarizations}

We first consider a laser pulse in the linear or circular polarization state interacts with the atomic rubidium. Suppose the polarized electric field is given by 
\begin{equation}
\mathbf{E}(t)=\hat{\epsilon}\mathcal{E}(t)\cos(\omega t + \phi),
\end{equation}
where the polarization vector $\hat{\epsilon} = \hat{z}$ or $(\hat{x}\pm i\hat{y})/\sqrt{2}$ for the linear or circular polarization, respectively. The D1 transitions is dictated by the selection rule of $\Delta m=0$ or $\pm 1$ ($q=\Delta m$), so, for each $m$, the Hamiltonian can be written in the four-state basis $\{|F=2, m\rangle, |F=3, m\rangle, |F'=2, m\pm q\rangle, |F'=3, m\pm q \rangle \}$ as
\begin{equation}
H(t)=   { 
\begin{pmatrix} 
0 & 0 & {\mu}_{22'}   E(t) &{\mu}_{23'}   E(t)  \\ 
0 & \delta & {\mu}_{32'} E(t) & {\mu}_{33'} E(t) \\
{\mu}^*_{22'} E(t)^* & {\mu}^*_{32'} E(t)^* & \omega_0 & 0 \\
{\mu}^*_{23'} E(t)^* & {\mu}^*_{33'} E(t)^* & 0 & \omega_0+\delta' 
\end{pmatrix} },
\end{equation}
where $\delta=\omega_{F=3}-\omega_{F=2}$, $\delta'=\omega_{F'=3}-\omega_{F'=2}$,  $\omega_0=\omega_{F'=2}-\omega_{F=2}$, and the dipole moments are given by $\mu_{FF'}= \langle F, m|er_q| F', m-q\rangle $. 

Since we consider ultrafast laser interactions, the interaction time is extremely short compared to the inverse of any hyperfine energy splitting, so the effective Hamiltonian under a resonant approximation~\cite{ScullyBook} can be alternatively written in the atomic frame as
\begin{equation}
H(t)=  
\begin{pmatrix}
0 & 0 & A & B  \\ 
0 & 0 & C & D \\
A^* & C^* & 0 & 0 \\
B^* & D^* & 0 & 0 
\end{pmatrix},
\label{Int_H}
\end{equation}
where $A=\mu_{22'}\mathcal{E}(t)/2$, $B= \mu_{23'}\mathcal{E}(t)/2$, $C= \mu_{32'}\mathcal{E}(t)/2$, and $D= \mu_{33'}\mathcal{E}(t)/2$. When $\phi=0 $ is assumed without loss of generality, the coupling matrix $V(t)$ is given by
\begin{equation}
V(t)=
\begin{pmatrix}
A & B  \\ 
C & D \\
\end{pmatrix},
\end{equation}
and the MS transformation decouples the system into two independent two-state systems through diagonalizing the matrices $V(t)V^\dagger (t)$ and $V^\dagger(t)V(t)$ that are given by
\begin{equation}
V(t)V^\dagger(t)=V^\dagger(t)V(t)=
\begin{pmatrix}
A^2+B^2 & AC+BD  \\ 
AC+BD & C^2+D^2 \\
\end{pmatrix}.
\end{equation}
Then, the coupling strength of each two-state system, or the square of each Rabi frequency, is determined by the eigenvalues~\cite{ShoreJMP2013} given by
\begin{widetext}
\begin{equation}
\lambda_{\pm}=\frac{A^2+B^2+C^2+D^2\pm\sqrt{(A^2+B^2+C^2+D^2)^2-4(AD-BC)^2}}{2}.
\label{eigenvalue}
\end{equation}
\end{widetext}
In the basis of the eigenvectors $\left|\lambda_\pm\right>$, an arbitrary ground state can be a pure superposition state or a mixed state of $\left|\lambda_\pm\right>$. If these two eigenvalues $\lambda_{\pm}$ are the same, the two independent oscillators represented by $\left|\lambda_\pm\right>$ vibrate coherently with the same Rabi frequency, thus the CPT could occur. Here, the condition for $\lambda_+=\lambda_-$ is given by
\begin{equation}
\{(A+D)^2+(B-C)^2\}\{(A-D)^2+(B+C)^2\}=0.
\end{equation}

As for the $\pi$ transitions ($q=0$), Clebsch-Gordon coefficient symmetries~\cite{angular momentum} allow $\langle 2m|z|2m'\rangle=-\langle 3m|z|3m'\rangle$ ($A=-D$) and  $\langle 2m|z|3m'\rangle = \langle 3m|z|2m'\rangle$ ($B=C$) for all $m=m'$, which satisfies $\lambda_+=\lambda_-$ and the Rabi frequencies are the same for all $m$'s. Therefore, the CPT of the total system occurs and, moreover, the system initially in any coherent-superposition or mixed ground-state simply undergoes simply a Rabi oscillation with the Rabi frequency given by 
\begin{equation}
\Omega(t)=\sqrt{\mu_{22'}^2+\mu_{23'}^2}\mathcal{E}(t)/2\hbar. 
\end{equation}
The Rabi oscillations in these complex degenerate-level manifolds have been observed in an implementation of single qubit operations of atomic rubidium~\cite{LimSR2014}. 

Meanwhile, for the $\sigma$ transitions  ($q=\pm1$), the equality $AD=BC$ holds, which makes one of the eigenvalues zero and the other twice the eigenvalue of $q=0$ (the $\pi$ transition). In other words, the $\sigma$ transitions for different $m$'s have respectively a pair of uncoupled states and their Rabi frequencies are $\sqrt{2}$ times bigger than that of the corresponding $\pi$ transition. Therefore, the temporal dynamics of $\sigma^\pm$ transitions depend on the initial condition, unlike the $\pi$ transitions, since its oscillators are drastically different, which behavior has been applied to coherent controls of medium gains~\cite{Delagnes}.

\subsection{Elliptic polarizations}
\label{ThyEP}
Now we generalize the problem by considering elliptically polarizations of the laser interaction~\cite{VitanovPRA2003}. When the electric field is defined by
\begin{equation}
\mathbf{E}(t)=\hat{\sigma}_+\mathcal{E}_+(t)\cos(\omega t + \phi_+)+\hat{\sigma}_-\mathcal{E}_-(t)\cos(\omega t + \phi_-),
\end{equation}
the elements in $V$ are given by $\mathcal{E}_\pm(t)\eta_{m}^{m'}e^{i\phi_\pm}/2$, where $\eta_{m}^{m'}$ are the transition dipole moment, $\langle F m | er_{\pm 1} | F' m' \rangle$. The dipole moments can be factorized by a common factor, $\mu_{JJ'}=\langle J=1/2 || er || J'=1/2\rangle$ for all $\{F, m\}$~\cite{angular momentum}. When the ellipticity $\epsilon$ is defined by $(\mathcal{E}_+^2-\mathcal{E}_-^2)/(\mathcal{E}_+^2+\mathcal{E}_-^2)$, the Rabi frequencies are given by
\begin{equation}
\Omega_\pm =\pm\Omega_o\sqrt{\frac{1\pm\epsilon}{3}}
\label{Rabipm}
\end{equation}
with $\Omega_o=\mu_{JJ'} (\mathcal{E}_+^2+\mathcal{E}_-^2)^{1/2}/{2\hbar}$.

When the Hamiltonian is MS-transformed through the summarized procedure in Eqs.~(1)-(4),  the resulting Hamiltonian $H'$ is a direct sum of twelve sets of two-state systems given by
\begin{eqnarray}
H' &= & \hbar 
\begin{pmatrix}
0 & \Omega_+e^{i\phi_+}  \\
\Omega_+e^{-i\phi_+} & 0 
\end{pmatrix}^{ \oplus 6}   \nonumber \\
& \oplus &
\begin{pmatrix}
0 & \Omega_-e^{i\phi_-}  \\
\Omega_-e^{-i\phi_-} & 0 
\end{pmatrix}^{ \oplus 6},
\label{H'}
\end{eqnarray}
where the first set of six two-state systems corresponds to the $\sigma^+$ transitions and the second to the $\sigma^-$ transitions. This result can be readily understood in terms of the basis reduction from  $|F, m_F\rangle$ to $|J, m_J\rangle$~\cite{PeggOC}. Since spin interactions (hyperfine interactions in this case) are irrelevant in ultrafast-time scale dynamics, they can be traced out, and the reduced basis description in $|J, m_J\rangle$ can be alternatively acquired. The advantage for using the MS-transformed $|F, m_F\rangle$ bases, instead of using the reduced basis set, comes from the fact that the coherent dynamics coupled with spin interactions can be directly seen in the former bases, implying applications to ultrafast time-scale nuclear-spin polarization controls~\cite{NakajimaPRL2007, HaydenPRL2008, NakajimaPRL2008}.

\section{Experimental Procedure}
\begin{figure}[htb]
\centering
\includegraphics[width=0.45\textwidth]{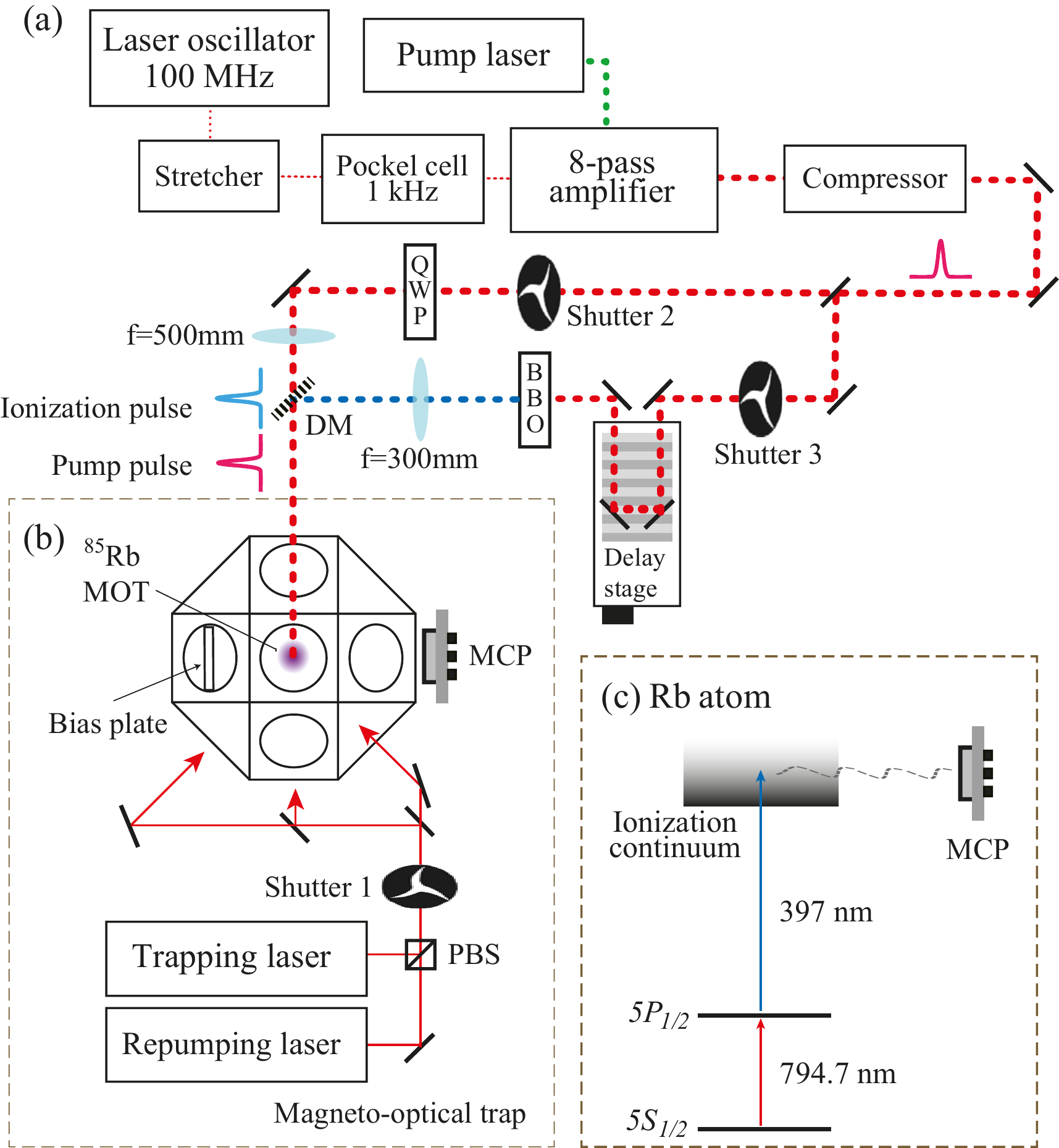}
\caption{(Color online) (a) Schematic diagram of the experimental setup. (b) Magneto-optical trapping apparatus for $^{85}$Rb. (c) Energy level diagram of $^{85}$Rb atoms.}
\label{FigA2}
\end{figure}

The experimental setup is schematically shown in Fig.~\ref{FigA2}. Rubidium atoms ($^{85}$Rb) were cooled and trapped in a conventional magneto-optical trap~\cite{PhillipsRMP1998}. The detail of the experimental setup was reported elsewhere~\cite{SangkyungPRA2012, LeeOL2015, LimSR2014}. Then the atoms were interacted with ultrashort laser pulses from a Ti:sapphire mode-locked laser amplifier operated at a pulse repetition rate of 1~kHz. The center wavelength of the laser pulses (the pump pulse) was tuned to 794.7~nm (or 780 nm) for the rubidium D1 (D2) transition with a full-width-half-maximum (FHWM) bandwidth of 3~nm (0.5~nm). The single pulse energy was varied up to 0.03~mJ, which was equivalent to a Rabi pulse area of $\Theta=2\pi$ when the pulse diameter at the atom was about 0.5~mm and the size of the atom cloud was 200~$\mu$m. The polarization was controlled by a broadband quarter-wave plate (QWP). 

The measurement of the excited state ($5P_{1/2}$ or $5P_{3/2}$) population of the atoms was achieved by photo-ionizing the excited atoms by frequency-doubled ultrafast pulses (the probe pulse) from the same laser. The pump and probe pulses were independently focused and delivered by a dichroic mirror (DM) to the MOT. The number of ions (Rb$^+$) was counted by a micro-channel plate detector (MCP). The signal linearity was assured by operating the experiment in the one-photon perturbation regime. The error due to three-photon ionizations by the pump pulse was estimated below 5\% at  a pulse-area of $3\pi$~\cite{LeeOL2015}, which was negligible in the considered experimental condition. The whole experiment (four steps: MOT-turn off, laser-control, ionization, MOT-turn on) was repeated at 2 Hz by cyclically turning on and off three mechanical shutters.

\section{Results and Discussion}
\label{SecResults}

\begin{figure}[h]
\centering
\includegraphics[width=0.45\textwidth]{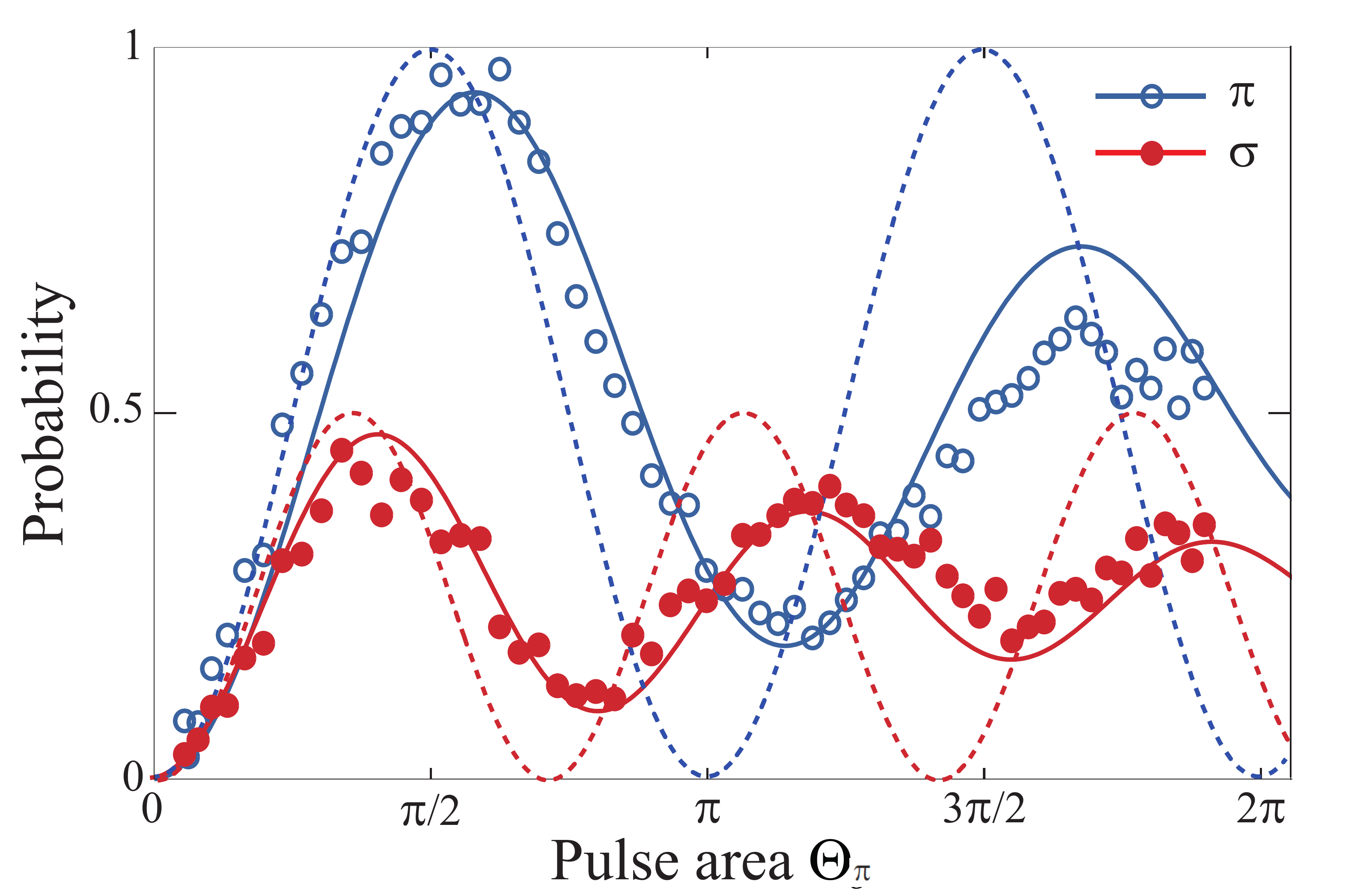}
\caption{(Color online) Ultrafast Rabi oscillations of $^{85}$Rb atoms between $5S_{1/2}$ and $5P_{1/2}$ energy states. Experimental results for linearly (the $\pi$ transition) and circularly (the $\sigma$ transition) polarized ultrafast laser pulses are shown with blue open circles and red filled circles, respectively. The two results are commonly scaled along the vertical axis by the maximal population of the $\pi$ transition and along the horizontal axis by the $\pi$-transition pulse area. Black dotted lines are the theoretical calculation based on Eq.~\eqref{P_elliptic} for the corresponding Rabi oscillations. New calculations that take the spatial average effect into account are represented in solid lines.}
\label{FigA3-2}
\end{figure}

The MS transformation of the rubidium D1 transitions, described in Sec.~\ref{SecTheory}, predicts the relation between the Rabi oscillations of the $\pi$ and $\sigma$ transitions given, in terms of the Rabi-oscillation pulse-areas, by
\begin{eqnarray}
\Theta_{\pi} &=& \frac{1}{\sqrt{2}}\Theta_{\sigma} , 
\label{rel1}
\end{eqnarray}
which relation is comes from Eq.~\eqref{Rabipm} that gives $\Omega(\epsilon=0)=\Omega_o/\sqrt{3}$ and $\Omega(\epsilon=\pm1)=\pm \sqrt{{2}/{3}} \Omega_o$ or zero. Also, with the same reason, a half of the ground levels remain intact for the $\sigma$ transitions if the atom is unpolarized, resulting in the relation of the maximal excitation probabilities given by
\begin{eqnarray}
{\rm P}^{\rm max}_{\pi} &=& 2 {\rm P}^{\rm max}_{\sigma}.
\label{rel2}
\end{eqnarray}
To verify these predictions, Rabi oscillation experiments of rubidium D1 transitions performed by $\pi$- and $\sigma$-polarized ultrafast laser pulses, respectively, were performed. The result is shown in Fig.~\ref{FigA3-2}, clearly exhibiting that the Rabi frequency of the $\sigma$ transition is $\sqrt{2}$ times larger than that of the the $\pi$ transition and that the oscillation amplitude of the $\sigma$ transition is a half of that of the the $\pi$ transition, thus the two relations in Eqs.~\eqref{rel1} and \eqref{rel2} are confirmed. The theoretical calculations are respectively plotted with dashed lines, of which the deviation from the experimental result is due to the spatially inhomogeneous laser interaction with the atom ensemble~\cite{LeeOL2015}. When the spatial averaging effect is taken into account (by assuming the Gaussian laser beam and atom ensemble size ratio is 2.5), the new calculations, shown with solid lines, agree well with the experimental results.

\begin{widetext}

\begin{figure}[h]
\centering
\includegraphics[width=0.8\textwidth]{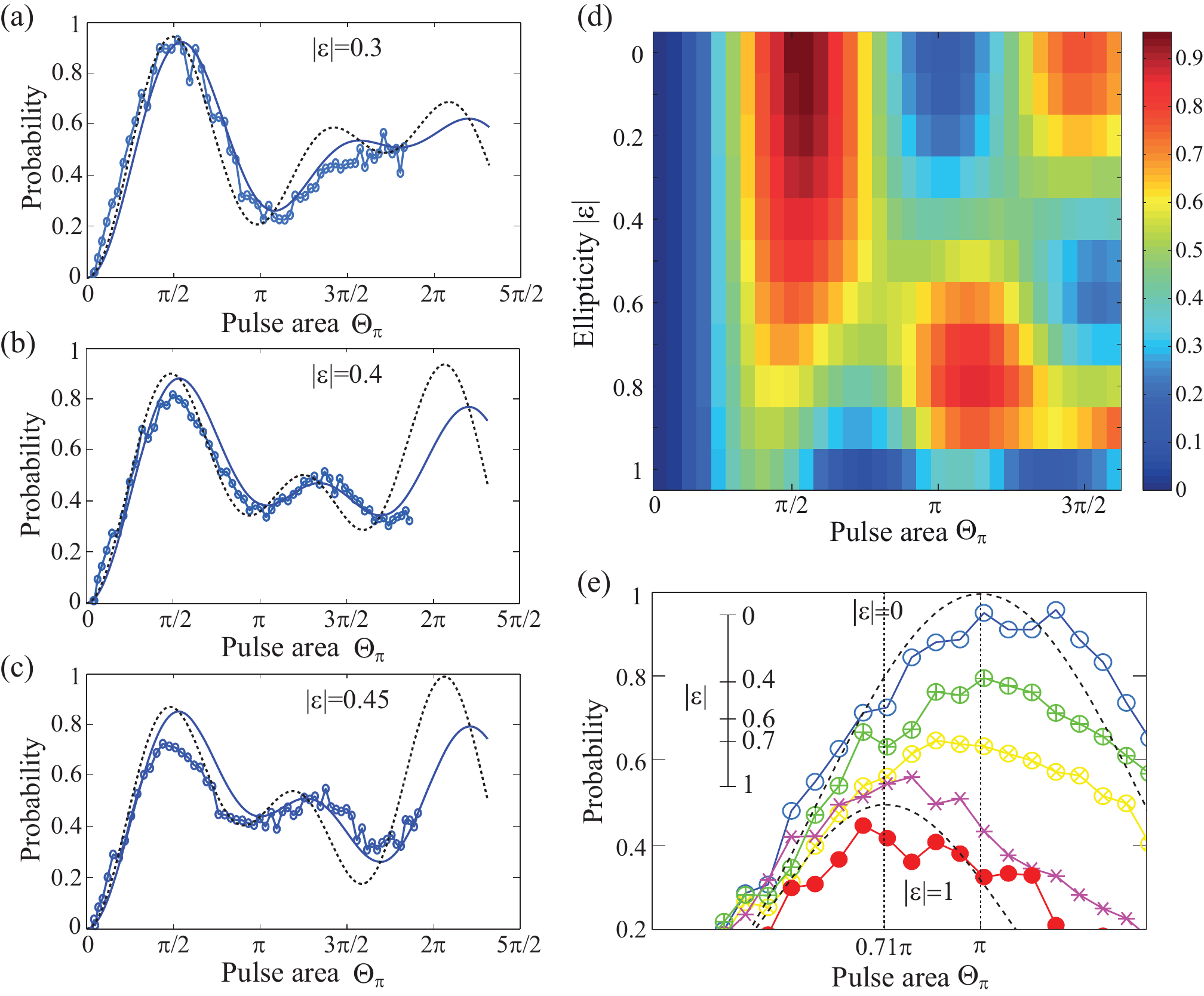}
\caption{(Color online) Experiment results for D1 transitions by elliptically polarized lights. (a-c) The blue circled lines represent experimental data for (a) $\epsilon=0.3$, (b)$\epsilon=0.4$, and (c)$\epsilon=0.45$ . The black dotted lines are the ideal Rabi oscillation curves and the blue solid lines are their spatially-averaged oscillations. (d) The calculated Rabi oscillations for various ellipticity are represented in a three-dimensional plot. (e) The experimental data of first peak position are closed up for various ellipticities.}
\label{Fig3}
\end{figure}
\end{widetext}

In the second experiment, elliptical polarizations are considered. When the rubidium atoms interact with an ultrafast laser pulse in an elliptical polarization, the interaction Hamiltonian in Eq.~\eqref{Int_H} is expressed as a superposition of $\sigma^+$ and $\sigma^-$ transitions. In this case, the decoupling is not simply attained through a 4-by-4 MS transformation $H'=UHU^\dagger$ in Eq.~\eqref{U}. However, we can still describe the dynamics of the MS-transformed two-state system as a mixture of the Rabi oscillations that correspond to the constituent polarization-specific transitions. In the D1 transitions performed by an elliptically polarized, there are parts of the two-state system that are respectively run by $\sigma^+$ and $\sigma^-$ transitions. In that regards, we can decompose the initial ground-state atom, written in $\left|J, m_J\right>$ bases, into $|\psi_\pm\rangle= a_\pm \left|1/2,\pm 1/2\right>$, where $|a_+|^2+|a_-|^2=1$, $|\psi_\pm\rangle$ are the Rabi oscillation bases, respectively, for $\sigma^\mp$ transitions. Therefore, the dynamics of the excited-state population of the atom under an elliptically polarized light is given by 
\begin{eqnarray}
P_{D1} &=& |a_+|^2  \sin^2{\Theta_+} + |a_-|^2 \sin^2{\Theta_-},
\label{P_elliptic}
\end{eqnarray}
where $\Theta_\pm=\sqrt{(1\pm \epsilon)/3}\Theta$ are the parametrized pulse areas responsible for the $\sigma^\pm$ transitions, respectively, obtained from Eq.~\eqref{H'}, with $\Theta =  (2\hbar)^{-1} \int_{-\infty}^{t} \left<1/2\left\|er\right\|1/2\right>\mathcal{E}(t)dt$. The $\pi$ transition is naturally defined for $\epsilon =0$.
Figure~\ref{Fig3} represents the experimental results. The excited state population in $5P_{1/2}$ appears as a degraded oscillation in each ellipticity $\epsilon\in \{0.3, 0.4, 0.45\}$ as shown in Figs.~\ref{Fig3}(a-c), because, as in Eq.~\eqref{P_elliptic}, it is given by a sum of two distinct oscillations. The calculation based on Eq.~\eqref{P_elliptic} and considering inhomogeneous pulse areas for various ellipticities is represented in Fig.~\ref{Fig3}(d). The Fig.~\ref{Fig3}(e) represents the experiment result that close up the first peak for various ellipticity. When the ellipticity is varied from zero to one, the first oscillation appears at a gradually smaller pulse area, and the pulse area for the first peak is changed from $\pi$ to $\pi/\sqrt{2}$, as shown in Fig.~\ref{Fig3}(d, e). 

\begin{figure}[htb]
\centering
\includegraphics[width=8cm]{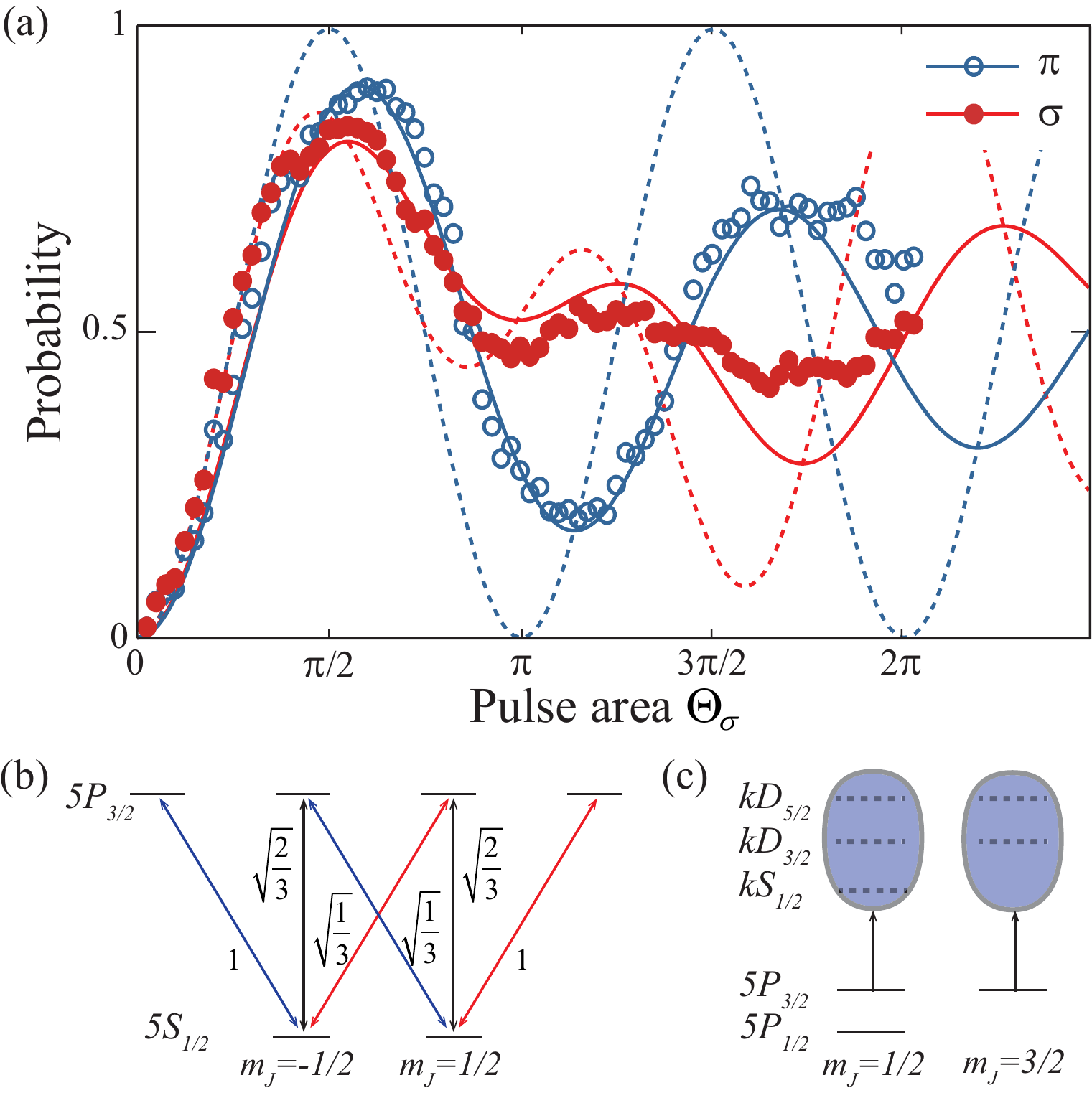}
\caption{(Color online) (a) Experiment results for D2 transitions. Experimental data are represented by blue open circles for circular polarizations and red closed circles for the linear polarization. In comparison, theoretical calculations and their spatially-averaged results are shown with dotted lines and solid lines, respectively. (b) Coupling strengths defined in $\left|J, m_J\right>$ bases. (c) Schematic comparison between the ionization schemes for $5P_{1/2}$ and $5P_{3/2}$ states. }
\label{Fig4}
\end{figure}

Lastly, Rabi oscillations of D2 transitions between $5S_{1/2}$ and $5P_{3/2}$ energy states are considered. 
In Fig.~\ref{Fig4}(a), the experimental results are compared with theoretical calculations. After a similar argument leading to Eq.~\eqref{P_elliptic}, the excited state population as a result of D2 transitions is also given by a sum of two Rabi oscillations as
\begin{eqnarray}
P_{D2} =|a_+|^2 \sin^2{\sqrt{\frac{2+\epsilon}{3}}\Theta}+ |a_-|^2  \sin^2{\sqrt{\frac{2-\epsilon}{3}}\Theta},
\label{D2P}
\end{eqnarray}
where $\Theta =  (2\hbar)^{-1} \int_{-\infty}^{t} \left<1/2\left\|er\right\|3/2\right>\mathcal{E}(t)dt$ and the $a_\pm$ denote the initial probability amplitudes of the ground state in $\left|J=1/2, m_J=\pm1/2 \right>$ bases. The Clebsch-Gordan coefficients for their coupling to $\left|J'=3/2, m_J'=\pm\right>$ bases are shown in Fig.~\ref{Fig4}(b).
However, a direct comparison of the excited-atom populations in $5P_{3/2}$ levels by counting photo-ions results in an experimental artifact, because the ionization passages via one-photon transitions to the continuum, as shown in Fig.~\ref{Fig4}(c), for $|J'=3/2, m_J'=\pm 1/2\rangle$ and $|J'=3/2, m_J'=\pm 3/2\rangle$, are different from each other. The ratio between the ionization rates may be estimated by comparing the bound-to-bound transitions of the corresponding angular symmetries. Since the coupling strengths from $|J=3/2, m_J=1/2\rangle$ and $|J=3/2, m_J=3/2\rangle$ are proportional to $\sum_{J'}\left| \langle 3/2 \left\| ez \right\| kJ'\rangle \left< kJ', 1/2|3/2, 1, 1/2, 0\right>\right|^2$ and $\sum_{J'}\left|\langle 3/2\left\| er \right\|kJ'\rangle\langle kJ', 3/2|3/2, 1, 3/2, 0\rangle \right|^2$, respectively, when $|\langle J\left\| er \right\|kJ'\rangle|^2\propto (2J'+1)/(2J+1)$ is assumed, it can be estimated that the ratio of the ionization rates is 1.06, which is in a good agreement with 1.1 used to fit the data to the theoretical calculations in Fig.~\ref{Fig4}(a). 

The remaining experimental errors are due to laser power fluctuations and the center mismatch between laser and atoms, both of which are predominant deviations at large pulse areas. The laser fluctuation within a 10\% of a shot-to-shot deviation was kept low by event statistics. The size of the atom cloud (200~$\mu$m) was small compared to the steering mirrors distance (300~$mm$), so the intrinsic misalignment of 30~$\mu$m due to steering angular resolution led to an experimental error. However, the guidelines that considered the spatial averaging effect are valid in the experiment~\cite{LeeOL2015}.

We now turn our attention to the implication of the obtained results to possible applications towards polarization controls of electron or nuclear spins.  Electron spin polarization controls by photo-excitation of atoms have been theoretically discussed for alkali atoms~\cite{Bouchene_19}, where coherent excitations to $nP$ states by circularly-polarized light was considered so that spin-orbit coupling exchanged spin and angular momenta. Alternatively, it can be considered, based on our demonstration, that the excited $nP$ state is independently controlled to $nP_{1/2}$ and $nP_{3/2}$ through ultrafast Rabi oscillations. So, for example, the procedure to control the electron polarization can be as follows: First, a circularly-polarized pulse resonantly tuned to $nP_{1/2}$ state completely excites the population in a particular electron spin state, for example $|m_j=1/2\rangle$, leaving the rest intact. The second pulse resonantly tuned to $nP_{3/2}$ state completely excites the remaining ground-state atoms, which are then photo-ionized by an extra means of an ultrafast time resolution. Finally the third pulse identical to the first pulse de-excites the atoms in $nP_{1/2}$ back to the ground state to complete the spin polarization control. Then, it is achieved that every ionized photo-electron has pure spin states. Higher fidelity controls may be achieved through adiabatic passages by applying frequency chirps.

\section{Conclusions}
In summary, we have studied ultrafast-laser driven Rabi oscillations of MS transformed atomic rubidium transitions. The complex coupling linkages involved with the D1 or D2 transitions are analyzed based on the MS transformation and the simplified decoupled systems are experimentally investigated by ultrafast laser interactions of spatially-localized rubidium atoms in a MOT in the strong interaction regime. It has been found that all the system dynamics are governed by a sum of at most two decoupled Rabi oscillations for any polarization state of the interacting ultrafast pulse. The implication of this result has been discussed to control photo-electron polarizations.


\end{document}